\documentclass[a4paper,12pt]{article}
\usepackage{pdflscape}
\usepackage{amsmath}
\usepackage{cite}
\usepackage{amssymb}
\usepackage{dcolumn}
\usepackage{bm}
\usepackage{color}
\usepackage{epsfig}
\usepackage{amsfonts}
\usepackage{graphicx}
\usepackage{subfigure}
\usepackage{dcolumn}
\usepackage{indentfirst}
\usepackage{authblk}
\usepackage{slashed}
\usepackage{appendix}
\usepackage{parskip}

\usepackage{booktabs}
\AtBeginDocument{
\heavyrulewidth=.08em
\lightrulewidth=.05em
\cmidrulewidth=.03em
\belowrulesep=.65ex
\belowbottomsep=0pt
\aboverulesep=.4ex
\abovetopsep=0pt
\cmidrulesep=\doublerulesep
\cmidrulekern=.5em
\defaultaddspace=.5em
}
\usepackage[colorlinks=false]{hyperref}
\hypersetup{citecolor=Blue}

\begin{document}

\vspace{80pt}

\centerline{\LARGE Holographic drag force with translational symmetry breaking }

\vspace{40pt}

\centerline{
 Sara Tahery, $^{\ast}$
\let\thefootnote\relax\footnote{$^{\ast}$  saratahery@htu.edu.cn}
Kazem Bitaghsir Fadafan $^{\dagger}$
\let\thefootnote\relax\footnote{$^{\dagger}$ bitaghsir@shahroodut.ac.ir} 
and Sahar Mojarrad Lamanjouei $^{\ddagger}$ 
\footnote{$^{\ddagger}$ s.mojarad.sm@gmail.com}}
\vspace{30pt}
{\centerline {$^{\ast}${\it Institute of Particle and Nuclear physics,  
Henan Normal University,  Xinxiang 453007, China
}}
\vspace{4pt} 
{\centerline {$^{\dagger,\ddagger}${\it Faculty of Physics, Shahrood University of Technology, P.O.Box 3619995161 Shahrood, Iran
}}
\vspace{40pt}

\begin{abstract}
In order to investigate how the drag force is affected by translational symmetry breaking (TSB), we  utilize a holographic model in which the background metric remains translational
symmetric while a graviton mass or other fields in the theory
break this symmetry. We calculate analytically the drag force, considering an asymptotic AdS$_5$ in which parameter $\alpha$  arises from TSB. This parameter can be intuitively understood as a measure of TSB strength and we anticipate that non-zero values of it will affect the drag force. In this asymptotic AdS$_5$ background, we will demonstrate that a decrease in $\alpha$ results in a reduction of the drag force. Moreover, we study the diffusion constant, which falls with increasing $\alpha$. It will eventually be shown that at lower values of $\alpha$ or $\mu$ (chemical potential), the transverse diffusion coefficient is larger than the longitudinal one, and the speed of the heavy quark has minimal impact on the ratio. 

\end{abstract}

\newpage

\tableofcontents

%\maketitle
\section{Introduction}
AdS/CFT conjecture originally relates the type IIB string theory on $AdS5 \times S_5$ space-time to the four-dimensional $\mathcal{N} = 4$ SYM gauge theory \cite{adscft}. In a holographic description of AdS/CFT, a strongly coupled field theory at the boundary of the AdS space is mapped to the weakly coupled gravitational theory in the bulk of AdS \cite{holo}. In this conjecture the  classical dynamics of a probe string  in the AdS space is taken into consideration on the gravitational side. 

The holographic drag force formalism is explained in \cite{drag,enlo,codr} as the momentum rate flowing down to the probe string is interpreted as the drag force exerted. In the gravitational side, the presence of probe branes in the AdS bulk breaks conformal symmetry and sets the energy scales so leads to corrections in $AdS_5$. Drag force was studied using Gubser's proposal in the context of non-conformal models in \cite{nakano2007,Talavera2007}.

In the context of the AdS/CFT, holographic lattices have been crafted using spatially dependent sources. These lattices demonstrate a finite DC conductivity that aligns with the results from ionic lattice calculations \cite{1308.0329,1201.3917}. It's important to note that the construction isn't limited to just explicit lattice structures. The phenomenon of momentum relaxation has been probed within the holographic domain as well. This is achieved by introducing neutral matter, serving as a vast sink where momentum from charge carriers can be transferred \cite{0705.3870}. These charge carriers are analyzed under the probe approximation \cite{0912.1061}. In an alternative approach, the study of momentum relaxation has been approached by adopting a theory that overtly disrupts diffeomorphism invariance within the bulk. This line of inquiry specifically involves the application of the non-linear massive gravity theory mentioned in \cite{1011.1232} which was introduced to the holographic framework by Vegh in \cite{Vegh2013}.

One may consider the Drude model in the condensed matter systems to study the motion of a probe heavy particle. It loses momentum and energy to the medium which leads to a drag phenomena. Two main mechanisms in a weakly coupled system are responsible for the drag force which are production of massless particles by bremsstrahlung and two body collisions. Study of the drag phenomena in the system helps to see which one is dominant in the system. For example radiation by a rotating heavy quark has been studied in \cite{0809.2869} and different channels have been discussed, the extensions have been done in \cite{1906.11621,Hou:2021own}. From the condensed matter point of view, the translational symmetry in the medium breaks either by introducing impurities or periodic potentials. As a result, one expects additional new mechanisms for the energy loss in the system. 

In this study, we investigate calculation of the drag force in systems where translational symmetry is not preserved. We employ  holographic models that address this symmetry breaking. Although these models maintain a symmetric background metric, the symmetry is interrupted by introducing either a mass for the graviton or other fields. For example, Vegh's model introduces mass terms for gravitons that interfere with the bulk diffeomorphisms, which correspond to translational symmetry in holography \cite{davison2013, Vegh2013}. Similarly, Andrade and Withers' approach uses massless scalar fields that exhibit a linear relationship with a spatial coordinate in the field theory \cite{Andrade2013}. A model with a massless two-form field that aligns linearly with a spatial coordinate in a conformal field theory (CFT) was introduced in \cite{Groz2018}. A comprehensive framework for translational symmetry breaking has been examined in \cite{1904.05785}, utilizing holography. This framework considers models of holographic massive gravity and manipulates translational symmetry, either explicitly or spontaneously, through a variable parameter. 

This paper is organized as follows, to use the TSB model in the drag force calculation, we briefly address it  in section \ref{TSBmodel}, after  computation of the drag force in such system in section \ref{Drag force} and studying the special case of $T=0$ in section \ref{zeroT} briefly, we discuss the diffusion coefficient in section \ref{Diffco} and Langevin coefficients in the TSB system in section \ref{flucTSB}. Summary will be covered in  section \ref{dis}.

%%%%%%%%%%%%%%%%%%%%%%%%%%%%%%%%%%%%%%%%%%%%%%%%%%%%%%%%%%%%%%%%%
\section{Translational symmetry breaking model} \label{TSBmodel}
%%%%%%%%%%%%%%%%%%%%%%%%%%%%%%%%%%%%%%%%%%%%%%%%%%%%%%%%%%%%%%%%%
A general representation of the translational symmetry breaking (TSB) model we would like to work with is as follows \cite{Andrade2013}, 
\begin{equation}\label{the model}
	S_0 = \int_M \sqrt{-g} \left[ R - 2 \Lambda - \frac{1}{2}\, \sum_{I}^{d-1} \,(\partial \psi_I)^2  - \frac{1}{4}\, F^2 \right ] d^{d+1} x 
	- 2 \int_{\partial M} \sqrt{-\gamma}\, K\, d^dx, 
\end{equation}
where  $R$ is  Ricci scalar, $\Lambda = - d(d-1)/(2 \ell^2)$, with AdS radius $\ell$, $F$ is the field strength $F=dA$ for a $U(1)$ gauge field $A$, and the $d-1$ massless scalar fields, $\psi_I$. The model is written  in the unit where $16\pi G=1$ and we will set $\ell=1$ also $\gamma$ is the induced metric on the boundary and $K$ is the trace of the extrinsic curvature, given by $K_{\mu\nu} = -\gamma_{\mu}^\rho\, \gamma_{\nu}^\sigma\,\nabla_{(\rho}n_{\sigma)}$ where $n$ is the outward pointing unit normal to the boundary. The  AdS$_{d+1}$ solutions are given by,
\begin{equation}\label{eq:metric}
ds^2= -f(r)dt^2 +\frac{dr^2}{f(r)}+ r^2\delta_{ab} dx^a\,dx^b,\quad A=A_t(r)dt. \quad \psi_{I}=\alpha_{I\,a}x^{a},
\end{equation}
where
 %$i = 1, 2, 3$ are orthogonal spatial boundary coordinates,
 $a$ labels the $d-1$ spatial $x^a$ directions, $I$ is an internal index that labels the
$d-1$ scalar fields and $\alpha_{I\,a}$ are real arbitrary constants. The ansatz 
Eq.\eqref{eq:metric} leads to the solution,
\begin{equation}\label{fz}
f(r)=r^2-\frac{\alpha^2}{2(d-2)}-\frac{m_0}{r^{d-2}}+\frac{\mu^{2}}{2}\frac{d-2}{d-1} (\frac{r_0}{r})^{2(d-2)}, 
\end{equation}
\begin{equation}\label{At}
A_t=\mu \Big{(} 1-\frac{r^{d-2}_0}{r^{d-2}}\Big{)},
\end{equation}
where,
\begin{equation}\label{alpha}
\alpha^2\equiv \frac{1}{d-1}\sum_{a=1}^{d-1} \vec{\alpha}_{a}.\vec{\alpha}_{a},
\end{equation}
in which,
\begin{equation}\label{alphadelta}
 \vec{\alpha}_{a}.\vec{\alpha}_{b}=\alpha^2\delta_{ab}\quad\quad\forall\,a,b.
\end{equation}
$m_0$ is proportional to the energy density of the brane and  can be acquired by solving  $f(r_0) = 0$ so that $r_0$ gives the horizon location,
\begin{equation}\label{m0}
m_0=r^d_0 \Big{(}1+\frac{d-2}{2(d-1)}\frac{\mu^2}{r^2_0}-\frac{1}{2(d-2)}\frac{\alpha^2}{r^2_0}\Big{)}.
\end{equation}
The temperature of the black hole is given by,
\begin{equation}\label{T}
T=\frac{f'(r_0)}{4\pi}=\frac{1}{4\pi}\Big{(}dr_0-\frac{\alpha^2}{2\,r_0}-\frac{(d-2)^2\,\mu^2}{2(d-1)r_0}\Big{)}.
\end{equation}
As our current work aims to work on an asymptomatic $AdS_5$, let's assume $d=4$ in the model Eq.\eqref{the model} which is written as, 
\begin{equation}\label{eq:metricd}
ds^2= -f(r)dt^2 +\frac{dr^2}{f(r)}+ r^2 dx_i^2,
\end{equation}
for $i=1,2,3$.
Also Eq.\eqref{fz}, Eq.\eqref{m0} and Eq.\eqref{T} are written as,
\begin{equation}\label{fzd}
f(r)=r^2-\frac{\alpha^2}{4}-\frac{m_0}{r^2}+\frac{\mu^{2}}{3}\frac{r^4_0}{r^4},
\end{equation}
\begin{equation}\label{m0d}
m_0=r^4_0 \Big{(}1+\frac{\mu^2}{3r^2_0}-\frac{\alpha^2}{4r^2_0}\Big{)},
\end{equation}
and,
\begin{equation}\label{Td}
T=\frac{1}{4\pi}\Big{(}4r_0-\frac{\alpha^2}{2\,r_0}-\frac{2\,\mu^2}{3r_0}\Big{)},
\end{equation}
respectively. As demonstrated by Eq.\eqref{Td},  since the temperature varies with $\alpha$ and $\mu$, it is not fixed, (therefore, when examining the drag force in the next section, we will make the quantities dimensionless).\\ 
Plugging Eq.\eqref{m0d} into Eq.\eqref{fzd} provides,
\begin{equation}\label{fm0cancelled}
f(r)=r^2\big{(}1-\frac{r^4_0}{r^4}\big{)}-\frac{\alpha^2}{4}\big{(}1-\frac{r^2_0}{r^2}\big{)}-\frac{\mu^{2}}{3}\frac{r^2_0}{r^2} \big{(}1-\frac{r^2_0}{r^2}\big{)}.
\end{equation}
Demanding that $T\geq 0$ gives us the constraint,
\begin{equation}\label{constariantT}
24 r^2_0-3\alpha^2-4\mu^2\geq 0,
\end{equation}
and  imposing the null energy
condition on the Ricci tensor of the metric function in Eq.\eqref{eq:metric}
gives,
\begin{equation}\label{nullenergycon2}
\alpha^2 +\frac{4\, r_0^4 \,\mu^2}{r^4}\geq 0.
\end{equation}
Therefore, in order to perform the calculations in the following sections, conditions Eq.\eqref{constariantT} and Eq.\eqref{nullenergycon2} must be met.

In the case of $r_0^2=\frac{\alpha^2}{8}+\frac{\mu^2}{6}$ the temperature vanishes. At $T=0$ the black hole becomes a finite entropy domain wall which interpolates between unit-radius AdS$_{d+1}$ in the UV and a near horizon AdS$_2 \times \mathbb{R}^{d-1}$, where the AdS$_2$ radius, $\ell_{\text{AdS}_2}$, is given by \cite{Andrade2013},
\begin{equation}\label{tzerodiscussion}
\ell_{\text{AdS}_2}^2 =\frac{1}{d(d-1)} \frac{(d-1)\alpha^2 + (d-2)^2\mu^2}{\alpha^2 + (d-2)^2\mu^2}.
\end{equation}
Note that this near horizon geometry can be obtained even when $\mu=0$, supported by $\alpha$.
%%%%%%%%%%%%%%%%%%%%%%%%%%%%%%%%%%%%%%%%%%%%%%%%%%%%%%%%%%%% 
\section{Drag force in a TSB system} {\label{Drag force}}
%%%%%%%%%%%%%%%%%%%%%%%%%%%%%%%%%%%%%%%%%%%%%%%%%%%%%%%%%%%%
 When a quark moves through a hot medium it quenches strongly. Drag force is related to the energy loss of heavy quark, based on the interaction between the
 quark and the medium. From string theory point of view the energy loss of the quark is understood as the momentum flow along a moving classical string into the horizon. In this conjecture the moving quark is mapped onto a probe string in the AdS space. So instead of studying the motion of quark  in a strongly coupled system one can simply consider the classical dynamics of a string on the gravitational side. Since $\mathcal{N} = 4$ SYM theory can capture most of the dynamics of  QGP  after a critical temperature, AdS/QCD approximate the phenomenology of QCD by replacing it with $\mathcal{N} = 4$ SYM theory. According to the   holographic drag force formalism momentum rate flowing down to the probe string is interpreted as the drag force  and the quark is prescribed to move on the boundary of $AdS_5$.  The bottom-up approach starts with a five-dimensional effective field theory somehow motivated by string theory and tries to fit it to QCD as much as possible (as examples see  \cite{zhang2018,xiong2019,sara2004}), also in \cite{Chen2024}  transport properties of Quark-Gluon Plasma with a machine-learning assisted holographic approach is studied.  
In the linear-axion model, momentum relaxation is achieved by introducing a linear axion field, leading to a drag force that is proportional to the velocity of the probe particle \cite{Chernicoff2012}. This results in a linear drag force behavior. The model exhibits a transition between coherent and incoherent metal phases, which affects the drag force. Additionally, the longitudinal shear viscosity in this model can violate the Kovtun-Son-Starinets (KSS) bound, indicating strong anisotropic effects. Similarly, in the massive gravity model, the introduction of massive gravitons breaks the diffeomorphism invariance, resulting in momentum dissipation and a non-zero drag force even at zero temperature \cite{1504.07635}. The drag force in massive gravity models is often linked to universal conductivity properties, with the presence of massive gravitons leading to a finite DC conductivity that affects the drag force behavior. Like the linear-axion model, the massive gravity model also shows phase transitions that impact the drag force, which can be studied through the behavior of the probe particle in the holographic setup. These features highlight some universal mechanisms, such as the role of momentum dissipation on drag force behavior.
 
We present the holographic calculation of the drag force in the presence of translational symmetry breaking using the model we previously introduced. 
 In order to compute the drag force, consider the Nambu-Goto action as,
\begin{equation}\label{eq:NG action}
S=-\frac{1}{2\pi\alpha'} \int d^2 \sigma \sqrt{-\det  g_{\alpha\beta} }\quad\quad g_{\alpha\beta}=G_{\mu \nu} \partial X_{\alpha}^{\mu} \partial X_{\beta}^{\nu},
\end{equation}
where $\sigma^{\alpha}$ are coordinates  of the string worldsheet and  $G_{\mu\nu}$ is the five-dimensional Einstein metric. To describe the late-time behavior of a string attached to a quark moving  in the $X_1$ direction in the thermal plasma with speed  $v$,
we write an ansatz that ought to meet the assumption that  the steady state behavior is obtained at late time. Therefore,   
\begin{eqnarray}\label{eq:x3}
X_0&=&t,\nonumber\\
X_1(r,t)&=&vt+\xi(r)+ \mathcal{O}(t),\nonumber\\
X_2&=&0,\nonumber\\
X_3&=&0, \nonumber\\
r&=&\sigma,
\end{eqnarray}
where
 $\mathcal{O}(t)$ are all terms that vanish at late time, from this point forward we ignore them.
From Eq.\eqref{eq:metricd} and Eq.\eqref{eq:x3} we find the Lagrangian as,
\begin{equation}\label{eq:Lagrangy}
\mathcal{L}=\sqrt{ 1-\frac{r^2}{f(r)} v^2+f(r) r^2 \xi'^2(r)},
\end{equation}
and from Eq.\eqref{eq:Lagrangy} the energy-momentum current is written as,
\begin{equation}\label{eq:pi1}
\Pi_{1}=\frac{\partial \mathcal{L}}{\partial \xi'}=\frac{\xi'(r)\,r^2\,f(r)}{\sqrt{ 1-\frac{r^2}{f(r)} v^2+f(r)\, r^2 \xi'^2(r)}},
\end{equation}
where $'$ denotes the derivation  with respect to $r$ in this case.
From Eq.\eqref{eq:pi1} one can obtain, 
\begin{equation}\label{epsilon}
\xi'(r)= \frac{\Pi_{1}}{r\,f(r)}\sqrt{\frac{f(r)-r^2\, v^2}{r^2\,f(r)-\Pi^2_{1}}}.
\end{equation}
The right side of  Eq.\eqref{epsilon} needs to be real in order to prevent an imaginary string ansatz. Thus there should be a common root between the numerator and denominator. To apply this condition first we find the root of numerator as, 
\begin{equation}\label{nominatorroot}
f(r_c)-r_c^2\, v^2=0,
\end{equation}
where $r_c$ is the root of Eq.\eqref{nominatorroot}. We define the dimensionless variables $\Tilde\alpha=\frac{\alpha}{2\pi T}$, $\Tilde\mu=\frac{\mu}{2\pi T}$, $\Tilde r=\frac{r}{2\pi T}$, $\Tilde r_0=\frac{r_0}{2\pi T}$  in order to solve the above equation. The Eq.\eqref{nominatorroot} is then expanded and written using the new dimensionless variables. We can eliminate the tilde sign in the following equation for simplicity's sake.  
\begin{equation}\label{fvrelation}
r_c^6(1-v^2)-r_c^4\,\frac{\alpha^2}{4}-r_c^2\,(r^4_0-\frac{\alpha^2}{4}\,r^2_0+\frac{\mu^{2}}{3}\,r^2_0)+\frac{\mu^{2}}{3}\,r^4_0=0.                                                                                                                                                                                                                                                                                 
\end{equation}
By solving the equation mentioned above, we can determine that,
\begin{eqnarray}\label{rcresult}
r_c&=&\frac{1}{4\sqrt{3}} \Bigg{(} \Big{[}
2^{1/3}\, \alpha^4 +  2^{7/3}\, r_0^2\, (3\, \alpha^2 - 4\, (\mu^2 + 3 r_0^2)\,)\, (v^2-1) \nonumber\\
&+&4 \alpha^2\, \Big{(}   
\frac{\alpha^6}{32}+ \frac{3}{16} \alpha^2 r_0^2 \,(3\, \alpha^2 - 4\, (\mu^2 + 3\, r_0^2)\,)\, ( v^2-1) - 9 \mu^2 r_0^4\, ( v^2-1)^2 \nonumber\\
& +&\Big{\lbrace}4\,(-\frac{\alpha^4}{16} + \frac{r_0^2}{4} \, (-3\, \alpha^2 + 4\, (\mu^2 + 3 \,r_0^2)\,)\, ( v^2-1)\,)^{3}\nonumber\\
&+&\frac{(\alpha^6 + 
  6 \, \alpha^2\, r_0^2\, (3\, \alpha^2 - 4 \, (\mu^2 + 3\, r_0^2)\,) ( v^2-1) - 
  288\, \mu^2\, r_0^4\, (v^2-1)^2)^2}{1024} \Big{\rbrace}^{1/2}\Big{)}^{1/3}\nonumber\\
&+&2^{11/3}\, \Big{(}
\frac{\alpha^6}{32}+ \frac{3}{16} \alpha^2\, r_0^2 \,(3\, \alpha^2 - 4\, (\mu^2 + 3\, r_0^2)\,)\, ( v^2-1) - 9 \,\mu^2\, r_0^4\, ( v^2-1)^2 \nonumber\\
& +&\Big{\lbrace}4\,(-\frac{\alpha^4}{16} + \frac{1}{4} r_0^2 (-3\, \alpha^2 + 4 \,(\mu^2 + 3\, r_0^2)\,) (v^2-1)\,)^{3}\nonumber\\
&+&\frac{(\alpha^6 +  6\, \alpha^2\, r_0^2 \,(3 \,\alpha^2 - 4 \,(\mu^2 + 3\, r_0^2)\,)\, ( v^2-1) -  288\, \mu^2\, r_0^4 \,( v^2-1)^2\,)^2}{1024} \Big{\rbrace}^{1/2}
\Big{)}^{2/3}
  \Big{]} \Bigg{/} \nonumber\\
&\Big{[}&  
(1 - v^2)   \Big{(}
\frac{\alpha^6}{32}+ \frac{3}{16} \alpha^2 r_0^2\, (3 \,\alpha^2 - 4\, (\mu^2 + 3\, r_0^2)) ( v^2-1) - 9\, \mu^2 r_0^4 ( v^2-1)^2 \nonumber\\
&+&\Big{\lbrace}4\,(-\frac{\alpha^4}{16} + \frac{1}{4} r_0^2 (-3\, \alpha^2 + 4\, (\mu^2 + 3\, r_0^2)) ( v^2-1))^{3}\nonumber\\
&+&\frac{(\alpha^6 + 
  6\, \alpha^2\, r_0^2\, (3\, \alpha^2 - 4 \,(\mu^2 + 3\, r_0^2)\,)\, ( v^2-1) - 
  288\, \mu^2\, r_0^4\, ( v^2-1)^2\,)^2}{1024} \Big{\rbrace}^{1/2}
\Big{)}^{1/3}
\Big{]}  
  \Bigg {).}^{1/2}\nonumber\\
\end{eqnarray}
As previously stated, the denominator of Eq.\eqref{epsilon}  must be satisfied by the above $r_c$ Eq.\eqref{rcresult} as, 
\begin{equation}\label{denomepsilon}
r_c^2\,f(r_c)-\Pi^2_{1}=0,
\end{equation}
which implies,
\begin{equation}\label{Pi}
\Pi_{1}=r_c\,\sqrt{f(r_c)}.
\end{equation}
One can write from Eq.\eqref{fm0cancelled},
\begin{equation}\label{fm0cancelledrc}
\sqrt{f(r_c)}=\sqrt{r_c^2\big{(}1-\frac{r^4_0}{r_c^4}\big{)}-\frac{\alpha^2}{4}\big{(}1-\frac{r^2_0}{r_c^2}\big{)}-\frac{\mu^{2}}{3}\frac{r^2_0}{r_c^2} \big{(}1-\frac{r^2_0}{r_c^2}\big{)}}.
\end{equation}
The current density, however, is indicated as follows, 
\begin{equation}\label{curden}
\pi_{1}=-\frac{1}{2\pi \alpha'} \xi' \frac{G_{tt}\,G_{11}}{-g}.
\end{equation}
As a result, the drag force is represented as,
\begin{equation}\label{eq:F}
F_{drag}=\frac{dp_1}{dt}=\sqrt{-g}\,\pi_{1}=-\frac{\Pi_{1}}{2\pi \alpha'}=-\frac{r_c\,\sqrt{f(r_c)}}{2\pi \alpha'},
\end{equation}
where $r_c$ is written as Eq.\eqref{rcresult} and $\sqrt{f(r_c)}$ is found from Eq.\eqref{fm0cancelledrc}. Note that the drag force's negative sign originates from the force vector's direction, so in order  to study $F$ we ignore the $-$, also take the $2\pi \alpha'=1$.
We continue by examining the drag force's behavior in plots. Keep in mind that we must verify constraints Eq.\eqref{constariantT} and Eq.\eqref{nullenergycon2} before selecting different parameter values. Since Eq.\eqref{nullenergycon2} is obviously satisfied for all values, we only concentrate on Eq.\eqref{constariantT}.  
\begin{figure}[h!]
\begin{minipage}[c]{1\textwidth}
\tiny{(a)}\includegraphics[width=8cm,height=5cm,clip]{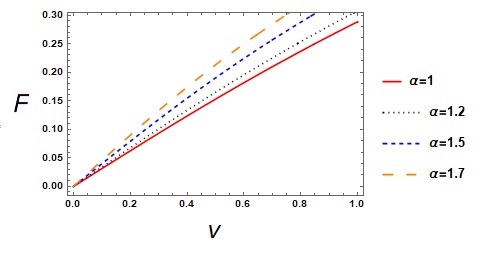}
%\hspace{0.2cm}
\tiny{(b)}\includegraphics[width=8cm,height=5cm,clip]{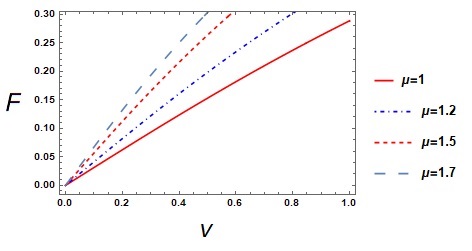}
\end{minipage}
\caption{Drag force versus speed $v$ for a) different values of $\alpha$ and $\mu=1$, b) different values of $\mu$ and $\alpha=1$. Plots are rescaled, but for the sake of simplicity, the quantities' tilde sign is removed.}
\label{fvsv}
\end{figure}
With $r_0=1$ set, figure \ref{fvsv} displays the drag force versus speed. Plot a) shows the effect of  $\alpha$, with $\mu=1$ fixed. The plot demonstrates that the drag force increases in magnitude as the TSB parameter increases. Likewise with fixed  $\alpha=1$  plot b) illustrates the effect of $\mu$, which increases the drag force.
Figure \ref{fvsalphamu} illustrates the drag force  with respect to a) chemical potential and b) parameter $\alpha$ for various speeds. According to both plots , the drag force is greater for larger values of $v$.
\begin{figure}[h!]
\begin{minipage}[c]{1\textwidth}
\tiny{(a)}\includegraphics[width=8cm,height=5cm,clip]{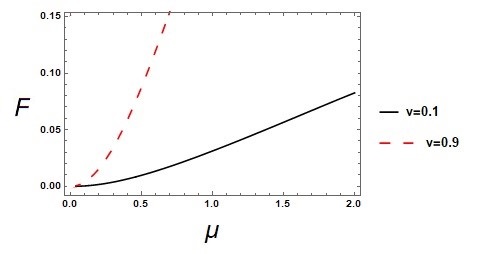}
%\hspace{0.2cm}
\tiny{(b)}\includegraphics[width=8cm,height=5cm,clip]{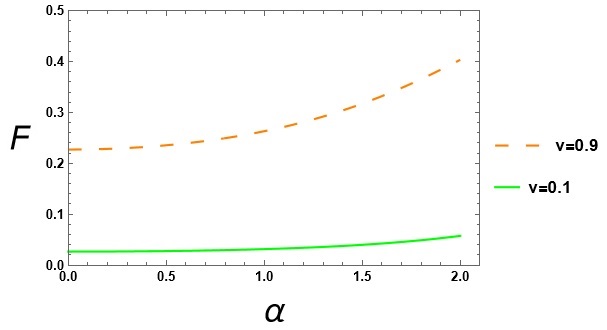}
\end{minipage}
\caption{Drag force  a) versus chemical potential $\mu$, b) versus $\alpha$ for different values of $v$. Plots are rescaled, but for the sake of simplicity, the quantities' tilde sign is removed. }
\label{fvsalphamu}
\end{figure}
\newpage
%%%%%%%%%%%%%%%%%%%%%%%%%%%%%%%%%%%%%%%%%%%%%%%%%%%%%%%%%
\section{Drag force at zero temperature in a TSB system}\label{zeroT}
%%%%%%%%%%%%%%%%%%%%%%%%%%%%%%%%%%%%%%%%%%%%%%%%%%%%%%%%%
That would be interesting to study motion of the heavy particle at zero temperature. One expects no energy loss in this case, however, we show that the particle experiences loosing momentum because of TSB. The drag force in extremal black holes at finite density has been studied in \cite{1512.05290}.

Examine constraint Eq.\eqref{constariantT} at precisely $T=0$, we may therefore discuss $F_{drag}$ in Eq.\eqref{eq:F} by determining different values of $(\alpha, \mu)$ at some fixed $r_0$.
\begin{figure}[h!]
\begin{minipage}[c]{1\textwidth}
\tiny{(a)}\includegraphics[width=8cm,height=5cm,clip]{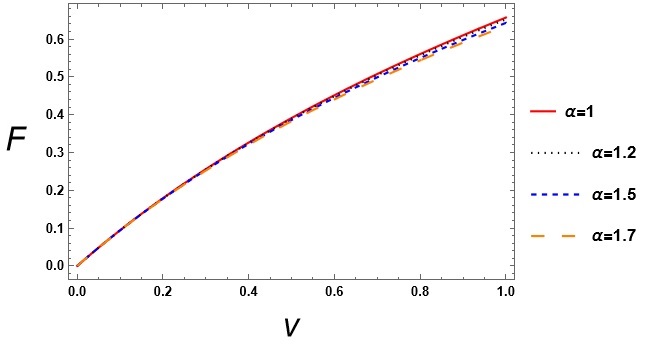}
\tiny{(b)}\includegraphics[width=8cm,height=5cm,clip]{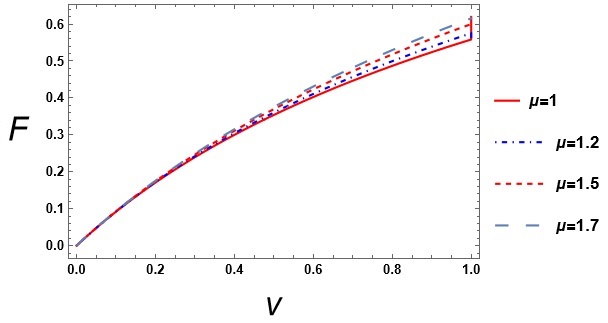}
\end{minipage}
\caption{ Drag force at zero temperature versus $v$ for a) different values of $\alpha$, b) different values of $\mu$. }
\label{FvsvzeroT}
\end{figure}
The drag force on a semi-classical string object moving at zero temperature is depicted in Figure \ref{FvsvzeroT}.  Plot (a) demonstrates that increasing $\alpha$ at zero temperature decreases the drag force  whereas plot  (b) indicates that increasing chemical potential $\mu$ has the opposite effect. We find the different effects of $\alpha$ in a zero temperature system and a thermal medium by comparing with figure \ref{fvsv}. 

Upon holding the system's parameters constant and varying the temperature, it is observed that the heavy particle is subjected to an increased drag force when the temperature is zero.
In terms of field theory, we observe that the entropy density remains non-zero even at zero temperature. This implies the existence of states to which a point particle can lose energy and momentum. Essentially, the system retains a non-zero density of states, allowing for the dissipation of momentum and energy by the point particle, even in the absence of thermal excitation. From a physical perspective, this non-zero entropy density at zero temperature indicates that there is a sufficient density of states available for a string to radiate momentum. This means that, despite the temperature being zero, the string can still interact with these states, losing momentum in the process. The presence of these states is crucial as it ensures that the string has a medium to transfer its momentum to, thereby maintaining the dynamics of the system even in a zero-temperature environment.
%\section{Drag force versus Temperature} 

Examining the behavior of the drag force versus temperature would be helpful, given the intriguing findings of studying drag force at finite temperature and zero temperature.  Keep in mind that rescaling has no point on this computation.  Thus, starting from \eqref{Td} we write in this manner,
\begin{equation}\label{Tr0}
T-\frac{1}{4\pi}\Big{(}4r_0-\frac{\alpha^2}{2\,r_0}-\frac{2\,\mu^2}{3r_0}\Big{)}=0, 
\end{equation}
subsequently $r_0$ as a function of $T$ will be obtained by solving \eqref{Tr0} for $r_0$. The drag force is expressed in terms of $\alpha$, $\mu$, and $T$ when we substitute $r_0$ in terms of T in \eqref{eq:F}.
\begin{figure}[h!]
\begin{center}$
\begin{array}{cccc}
\includegraphics[width=10cm]{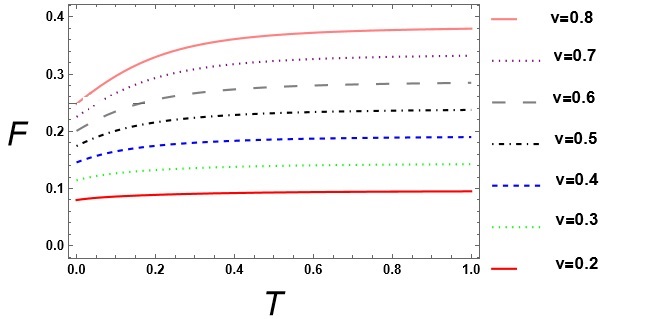}
\end{array}$
\end{center}
\caption{Drag force versus T for arbitrary fixed values of  $\alpha$ and  $\mu$ and different values of $v$.}
\label{FvsTv}
\end{figure}
The drag force versus temperature for various values of $v$ is displayed in figure \ref{FvsTv}. It is demonstrated that the drag force has an almost monotonous behavior with respect to temperature when $\alpha$ and $\mu$ are fixed. On the other hand, when increasing velocity is approaching, this behavior may gradually change. When the velocity is high enough, as in the highest $v$ plot, the significant difference between the drag force at low temperature and high temperature is clearly expressed, even though the $F$ appears to be a constant function at the lowest $v$ plot. 
\begin{figure}[h!]
\begin{minipage}[c]{1\textwidth}
\tiny{(a)}\includegraphics[width=8cm,height=5cm,clip]{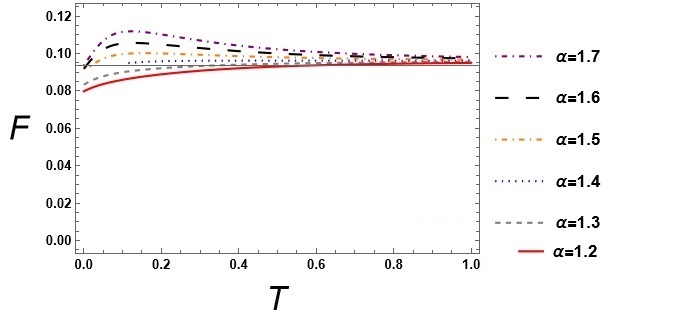}
\tiny{(b)}\includegraphics[width=8cm,height=5cm,clip]{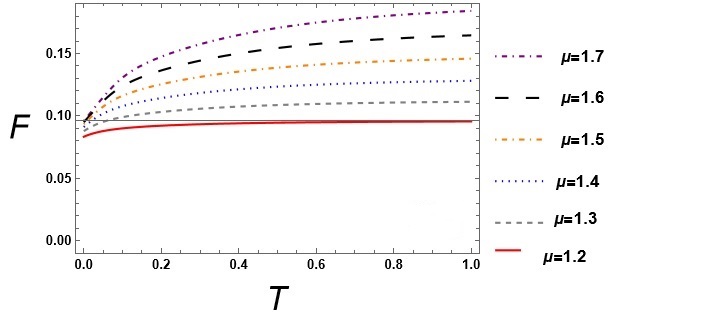}
\end{minipage}
\caption{ Drag force versus temperature for a) different values of $\alpha$ with arbitrary fixed values of $v$ and $\mu$ and  b) different values of $\mu$ with arbitrary fixed values of $v$ and $\alpha$}
\label{FvsT}
\end{figure}
Figure \ref{FvsT} plots also show F versus T in the following scenarios: a) $\mu$ and v are fixed but $\alpha$ is increasing; b) $\alpha$ and v are fixed but $\mu$ is increasing. Plot (a) demonstrates that the drag force develops a maximum value when $\alpha$ is increasing. This peak is apparent at low temperatures when $\alpha$ reaches a sufficient value, and at higher temperatures, the drag force plots with various $\alpha$ values are converging. Plot b) shows that, the drag force increases with increasing chemical potential as we knew from previous sections; however, there is no maximum $F$ when $\mu$ is increasing. 

Backreactions typically become significant at extremely low temperatures, which suggests that the probe limit of the probe string approach may encounter issues at zero temperature. In such conditions, the interactions between the probe string and the background cannot be ignored, as they can lead to substantial deviations from the expected behavior. This implies that the simplifications assumed in the probe limit might break down, necessitating a more comprehensive treatment that accounts for the backreactions to accurately describe the system's dynamics at zero temperature. Consequently, the reliability of the probe string approach in these extreme conditions is questionable, highlighting the need for alternative methods or adjustments to the existing framework to ensure accurate modeling.
\section{Diffusion Coefficient} \label{Diffco}
%%%%%%%%%%%%%%%%%%%%%%%%%%%%%%%%%%%%%%%%%%%%%%%
The diffusion coefficient, a fundamental parameter of plasma at RHIC and LHC for heavy quarks, which is related to the temperature, the heavy quark mass and the relaxation
time  is defined by  \cite{drag,enlo},
\begin{equation}\label{D}
D=\frac{T}{m} t_D, 
\end{equation}
where $T$ is the temperature, $m$ is the particle mass and $t_D$ is the damping time. It can be derived
from the drag force as,
\begin{equation}\label{tD}
F_{drag} \int _{0}^{t_D}dt=\int_{p}^{0}dp. 
\end{equation}
% Recall that $L^2=\sqrt{N}\, g_{_{YM}}\, \alpha'= \sqrt{\lambda}\, \alpha'$.
 % From Eq.\eqref{eq:F} and Eq.\eqref{tD} we obtain, 
%\begin{equation}\label{tDGC}
%t_D=,
%\end{equation}
Therefore from Eq.\eqref{D} the diffusion coefficient is written as,
\begin{equation}\label{DGC}
D=\frac{P}{m}\frac{T}{F}=T\frac{v}{\sqrt{1-v^2}}\frac{1}{F}.
\end{equation}
Where $F$ is written as Eq.\eqref{eq:F}. Like we did  for the drag force in the previous section, we examine how coefficient $D$ behaves in relation to  various (dimensionless) variables in the plots that follow.
\begin{figure}[h!]
\begin{minipage}[c]{1\textwidth}
\tiny{(a)}\includegraphics[width=8cm,height=5cm,clip]{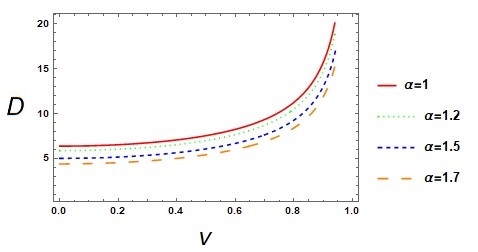}
%\hspace{0.2cm}
\tiny{(b)}\includegraphics[width=8cm,height=5cm,clip]{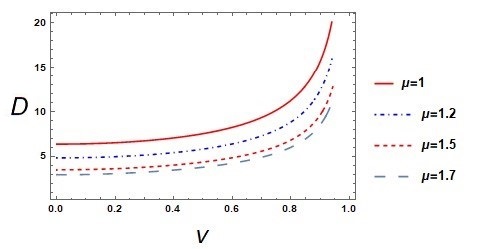}
\end{minipage}
\caption{Diffusion Coefficient versus speed,  for a) different values of $\alpha$ and $\mu=1$ , b) different values of $\mu$ and $\alpha=1$.  }
\label{Dvsv}
\end{figure}
Diffusion coefficient versus speed is plotted in figure \ref{Dvsv} for two scenarios: a) fixed $\mu$ and different values of $\alpha$, and b) fixed $\alpha$ and different values of $\mu$. In both scenarios, the diffusion coefficient decreases as the TSB parameter or chemical potential increases.\\ 
We represent the diffusion coefficient with respect to $\mu$ and $\alpha$ in figure \ref{Dvsalphamu}  for both low and high speeds, in order to examine how the diffusion coefficient varies with speed. 
The diffusion coefficient rises for high speed in both scenarios, but it changes noticeably when $D$ is versus $\alpha$.                                                                                                                                                                                                                                                                                                                                                                                                                                                                                                                                                                                                                                                                                                                                                                                                                                                                                                                                                                                                                                                                                                                                                                                                                                                                                                                                                                                                                                                                                                                                                                                                                                                                                                                                                                                                                                                                                                                                                                                                                                                                                                                                                                                                                                                                                                                                                                                                                                                                                                                                                                                                                                                                                                                                                                                                                                                                                                                                                                                                                                                                                                                                                                                                                                                                                                                                                                                                                                                                                                                                                                                                                                                                                                                                                                                                                                                                                                                                                                                                                                                                                                                                                                                                                                                                                                                                                                                                                                                                                                                                                                                                                                                                                                                                                                                                                                                                                                                                                                                                                                                                                                                                                                                                                                                                                                                                                                                                                                                                                                                                                                                                                                                                                                                                                                                                                                                                                                                                                                                                                                                                                                                                                                                                                                                                                                                                                                                                                                                                                                                                                                                                                                                                                                                                                                                                                                                                                                                                                                                                                                                                                                                                                                                                                                                                                                                                                                                                                                                                                                                                                                                                                                                                                                                                                                                                                                                                                                                                                                                                                                                                                                                                                                                                                                                                                                                                                                                                        \begin{figure}[h!]
\begin{minipage}[c]{1\textwidth}
\tiny{(a)}\includegraphics[width=8cm,height=5cm,clip]{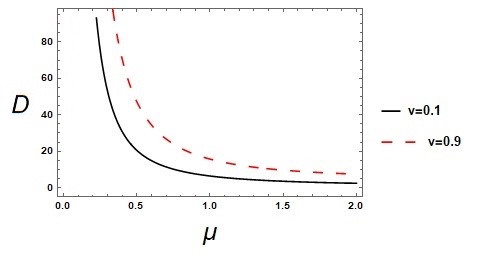}
\hspace{0.2cm}
\tiny{(b)}\includegraphics[width=8cm,height=5cm,clip]{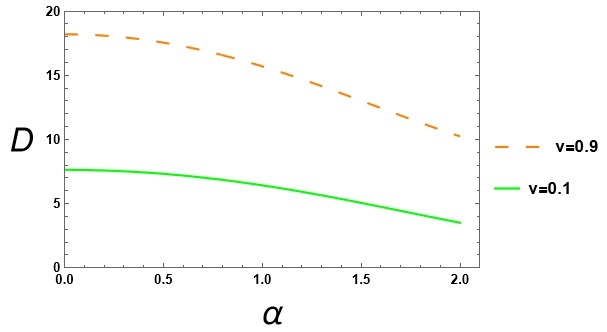}
\end{minipage}
\caption{Diffusion Coefficient  a)  versus chemical potential,  b)  versus $\alpha$ for different values of $v$.}
\label{Dvsalphamu}
\end{figure}

%\newpage
\section{Fluctuations on the string in translational breaking background}\label{flucTSB}

Heavy quarks out of equilibrium exhibit Brownian-like motion under the influence of a stochastic force, denoted as $\xi(t)$. These dynamics produce observables that are essential to comprehending the behavior of the quark gluon plasma. One can find a thorough synopsis of the physical principles in \cite{0903.1096}. In \cite{hep-ph/0605199,0812.5112,0903.1840} an  extensive research has been done on Langevin coefficients within the gauge/gravity duality. Also a universal methodology applicable to a broad spectrum of gravity theories, utilizing the membrane paradigm, was introduced in \cite{1310.6725}.

A quark moving with a constant velocity exhibits dynamics similar to Brownian motion. Its motion can be described using generalized Langevin equations, which incorporate the components of real-time correlation functions for the time-dependent $\xi(t)$. Assuming that, over long periods, the time-correlation functions are proportional to delta functions, the Langevin equations become local, and the diffusion coefficients remain constant.  In \cite{Giataganas:2013zaa}, it was argued that a universal inequality exists between transverse $\kappa_T$ and longitudinal $\kappa_L$ Langevin coefficients in generic isotropic backgrounds. It was shown that this inequality is only violated in anisotropic backgrounds, leading to negative excess noise in the system. In this work, we investigate the violation of this inequality in the background described by \eqref{eq:metric}. First we explain that the Browniaan's motion of the quark can be modeled by the generalized Langevin equations.  Two-point correlators for transverse and longitudinal to the direction of motion are given by the following equations
	\begin{eqnarray}\label{xiTL}
		\left< \xi_T(t) \xi_T(t') \right>=\kappa_T \delta(t-t'), \,\,\,\,\,
		\left< \xi_L(t) \xi_L(t') \right>=\kappa_L \delta(t-t'),
	\end{eqnarray}
	Where $\kappa_L$ and $\kappa_R$ are related to anti-symmetric retarded correlator $G_R$ as 
	\begin{equation}
		\kappa_T= -2 T_{ws} \lim_{\omega\rightarrow 0} \frac{Im G_R^T(\omega)}{\omega},\,\,\,\,\,\,	\kappa_L= -2 T_{ws} \lim_{\omega\rightarrow 0} \frac{Im G_R^L(\omega)}{\omega}
	\end{equation}
	In these relations the world sheet temperature is denoted by $T_{ws}$.

Next, we study the relativistic Langevin coefficients in the translational symmetry background Eq.\eqref{eq:metric} using holography. Our main goal is to study these coefficients. Here, we analyze the stochastic forces exerted on a quark that is in motion at a uniform speed. These forces arise due to the existence of a horizon on the string world-sheet, $r_c$. These equations incorporate Langevin coefficients that are proportional to $T_{ws}$, reflecting the thermal context of the motion \cite{Giataganas:2013zaa}. One finds the ratio of longitudinal to transverse Langevin coefficients as,
\begin{equation}\label{Kl/Kt}
	\frac{\kappa_L}{\kappa _T}= \frac{-3\alpha^2 r^4+24r^6 +4\mu^2r_c}{24r^2r_c^4 +3\alpha^2(r^4-2r^2r_c^2)+\mu^2(-8r^2r_c^2+12r_c^4)} 
\end{equation}
\begin{figure}[h!]
\begin{minipage}[c]{1\textwidth}
\tiny{(a)}\includegraphics[width=8cm,height=5cm,clip]{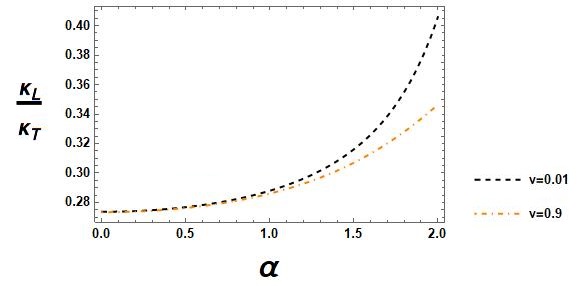}
%\hspace{0.2cm}
\tiny{(b)}\includegraphics[width=8cm,height=5cm,clip]{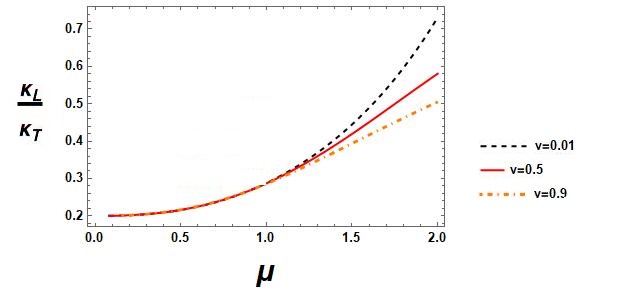}
\end{minipage}
\caption{Ratio of the  Langevin coefficients  a)  versus $\alpha$, b) versus $\mu$ for different $v$.}
\label{Ration}
\end{figure}
In Figure \ref{Ration}, we present a detailed plot illustrating the ratio as a function of the parameters $\alpha$ and $\mu$. The left-hand side of the figure is dedicated to the variation of the ratio with $\alpha$, while the right-hand side correlates the ratio with $\mu$. It is evident from the graph that when the ratio falls below one, the longitudinal diffusion coefficient is smaller than its transverse counterpart. This observation aligns with the arguments presented in reference \cite{1310.6725}, where it was anticipated due to the violation of translational symmetry, which in turn affects the proposed bound on the ratio. The parameters $\alpha$ and $\mu$ are chosen within a valid range that ensures the positivity of the temperature, a critical condition for the physical relevance of our model. The plots distinctly show that at lower values of $\alpha$ or $\mu$, the speed of the heavy quark has a negligible impact on the ratio. However, as $\alpha$ or $\mu$ increase, the influence of the quark's speed  becomes significantly more pronounced. This trend suggests that the speed of the heavy quark is a crucial factor in determining the behavior of the diffusion coefficients, especially in regimes of high $\alpha$ or $\mu$. The enhanced effect at larger values points to a non-linear dependency, which could be indicative of underlying complex dynamics that merit further investigation. Such dynamics may involve intricate interactions between the quark and the medium it traverses, potentially leading to novel insights into the nature of quark-gluon plasma and the fundamental principles governing it.
\section{Summary}\label{dis}
This paper investigated the effects of translational symmetry breaking on the drag force of a system. 
To do so, we used a holographic model in which the parameter $\alpha$ introduces the symmetry breaking. 
The effects of particle speed, chemical potential $\mu$, and non-zero values of $\alpha$ in such a system are investigated.
Plots revealed that as $\alpha$ or $\mu$ increase, the drag force likewise increases in magnitude, demonstrating that the strength of the TSB causes the drag force to increase. 
Furthermore, when plotted against $\alpha$ or $\mu$, we found that in both cases, the drag force increases with increasing particle speed, 
 the increase becomes statistically significant when plotted against $\alpha$. 
 At zero temperature, increasing chemical potential $\mu$ results in an increase in drag force; however, increasing $\alpha$ has the opposite effect.
 Consequently, we discover that $\alpha$ has different effects in a system running at zero temperature than it does in a thermal medium.
The examination also included analysis of the diffusion coefficient.
It turns out that when the TSB parameter or chemical potential increases, $D$ decreases.  Furthermore, in both scenarios, the diffusion coefficient increases for high velocities; however, it manifestly varies when $D$ is versus $\alpha$.
Finally, we discussed how the heavy quark's speed has very little effect on the ratio at lower values of $\alpha$ or $\mu$, and how the longitudinal diffusion coefficient is smaller than the transverse one.  Having said that, as $\alpha$ or $\mu$ rise, the influence of the quark's speed increases noticeably. In future works, it would be interesting to discuss the effect of TSB at high enough speed, where one can find the spatial string tension at non-zero chemical potential and the quark string configurations.

%\textbf{\Large{Acknowledgement}}\\

\end{document}